\documentclass[preprint,12pt]{elsarticle}

\usepackage{todonotes}
\usepackage{blkarray}
\usepackage[utf8]{inputenc}
\usepackage{graphicx}
\usepackage{amssymb}
\usepackage{amsmath}
\usepackage{empheq}
\usepackage{lineno}

\biboptions{comma, sort&compress, numbers, super}
\begin{document}

\begin{frontmatter}


\title{Modeling news spread as an SIR process over temporal networks}



\author[fgv]{Elisa Mussumeci}
\author[fgv]{Flávio Codeço Coelho}

\address[fgv]{Applied Mathematics School - Fundação Getulio Vargas}

\begin{abstract}
News spread in internet media outlets can be seen as a contagious process generating temporal networks representing the influence between published articles. In this article we propose a methodology based on the application of natural language analysis of the articles to reconstruct the spread network. From the reconstructed network, we show that the dynamics of the news spread can be approximated by a classical SIR epidemiological dynamics upon the network. From the results obtained we argue that the methodology proposed can be used to make predictions about media repercussion, and also to detect viral memes in news streams.
\end{abstract}

\begin{keyword}
News \sep SIR model \sep Epidemics \sep Temporal 
Networks


\end{keyword}

\end{frontmatter}

%\linenumbers

\section{Introduction}
\label{S:intro}
The internet is the main channel for dissemination of information in the 21st 
century. Information of any kind is posted online and is spread via 
recommendations or advertisement\cite{hermida2012share,romero2011influence}. 
Studying the dynamics of the spread of 
information through the internet, is a very relevant and challenging activity, 
since it can help the understanding of the factors which determine how far and 
how fast a given information can go. The most common way to observe information 
flow on the web, is by tracking how many times a given piece is replicated by 
different users over a period of time. Sometimes the content is modified as it 
is replicated, making it harder to track.

In the specific case of news articles, a number of factors influence their 
spread. Among the most important are the reputation of the original publisher 
-- though that is not easy to measure, and  the 
size of readership of a particular publisher, which will determine the initial 
spread of any piece. However the topology of the resulting network associated 
with dissemination of news cannot be anticipated and will depend on the 
subject of each news piece and its resonance with public interests.

In this work we decided to look at the spread of news stories over the internet 
characterizing the resulting spread network and the dynamics of the spread. We 
start by looking at an actual case of news spread, and estimate the spread 
network by applying ideas of temporal networks and topic Modeling, connecting 
similar articles within the bounds of temporal window of influence. Then we 
postulate that the spread dynamics approximates an epidemic process and model 
it 
using a Network SIR model\cite{pastor2015epidemic}. The spread of ideas as an 
epidemic process is not a new idea\cite{bettencourt2006power}, but here we 
Propose new tools to estimate the spread network from data and compare it with 
simulated networks produced by an SIR epidemic model.

\section{Methodology}
\label{S:methods}
\subsection{Data sources}
The data used for this study was obtained from the Media Cloud Brasil project 
(MCB)
which collects news articles from thousands of sources in the Brazilian 
Internet since 2013. From the MCB database we obtained 2129 articles talking 
about the Charlie Hebdo terrorist attack in February 2015. The articles span 
from the day of the attack to the end of march of 2015. The data include the 
full text of the article, the URL of publication and the date and time of the 
publication.

\begin{figure}[!ht]
 \centering
 \includegraphics[scale=0.8]{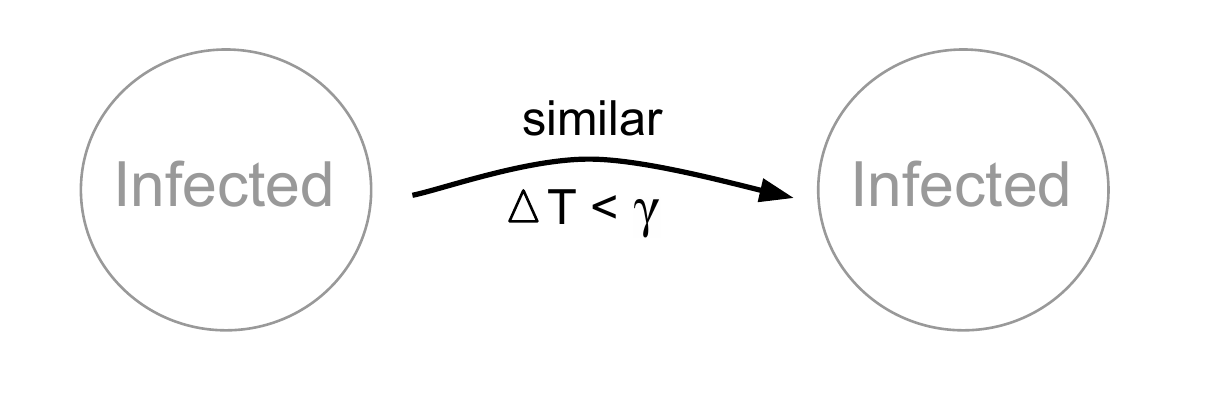}
 % nodes_graph.pdf: 348x118 pixel, 72dpi, 12.28x4.16 cm, bb=0 0 348 118
 \caption{Infection between articles}
 \label{fig:infection}
\end{figure}

\subsection{Article similarity}
In order to calculate a measure of similarity between text documents one can 
rely on a number of metrics for textual distance described in the 
literature\cite{mihalcea2006corpus}. Most of these metrics are based on a 
bag-of-words 
representation of texts, meaning that texts are defined in terms of which 
words they contain and their frequency in the documents. Such representation 
completely 
disregards higher level linguistic features of texts such as syntactics and 
semantics.  In 
this analysis, we want to use semantically similarity to describe the 
association between articles. In order for a news article to influence another, 
they must talk about the same concepts.

In order to capture the semantics of the articles we started by building a 
word vector representation for every word in our corpus' vocabulary, taking into 
account the coocurrence of words within a sentence. This 
model is built from a larger corpus of news articles (approximately 2.6 
million articles) according to the Skip-gram model, which has been shown to 
map the words to a vector space where semantic similarity is highly correlated 
with the cosine distance between word vectors~\cite{mikolov2013efficient}. This 
larger 
corpus corresponded to the total collection of article of the MCB project. The 
importance of training the word vector model on a corpus as large as 
possible, is that one gets a more accurate semantic representation of 
each word as a vector. It is important that the larger corpus represents a 
similar informational space as the sample we are trying to analyze.

\begin{table}
\begin{center}
\caption{Parameters of the skip-gram model.}
\label{tab:pars}
\begin{tabular}{lcc}
\textbf{Parameter} & \textbf{Value} & \textbf{Meaning} \\
\hline 
Minimum word count & 10 & word minimum frequency in the corpus.\\
Number of features & 300 & Dimension of word vectors\\
Context & 10 & Text window around word \\
\hline
\end{tabular}
\end{center}
\end{table}

The word vector model, was trained with the parameters described in table 
\ref{tab:pars}. The fitted word vector model consists of a matrix of $m$ word 
vectors ($w_i$) as rows. Each row represents am $n$-dimensional feature vector, 
with $n=300$:
\[
\begin{blockarray}{ccccc}Pm
& f_1 & f_2 & \hdots & f_n \\
\begin{block}{c[cccc]}
w_1 & a_{11} & a_{12} & \hdots & a_{1n} \\
w_2 & a_{21} & a_{22} & \hdots & a_{2n} \\
\vdots & \vdots & \vdots & \ddots & \vdots\\
w_m &  a_{m1} & a_{m2} & \hdots & a_{mn}\\
\end{block}
\end{blockarray}
\]

From the word vectors obtained, we created document vectors defined as a 
weighted sum of word vectors. For a document $d$ containing $k$ distinct words, 
its vector representation $\overrightarrow{D}$ is given by

\begin{equation}
\label{eq:doc_vector}
 \overrightarrow{D} = \sum_{i=1}^k w_i \times \mathcal{W}_{w,d}
\end{equation}

where $\mathcal{W}_{w,d}$ is the weight of the word $w_i$ in the document $d$. 
This weight can be calculated in different ways, for this work we used the 
TFIDF 
score~\citep{Hiemstra_2000} of the word in the document. Another possibility 
would be to use the 
frequency of the word in the document.

From the weighted sum we obtain document vectors which can be represented by 
the matrix below

\[
\begin{blockarray}{ccccc}
& f_1 & f_2 & \hdots & f_n \\
\begin{block}{c[cccc]}
d_1 & a_{11} & a_{12} & \hdots & a_{1n} \\
d_2 & a_{21} & a_{22} & \hdots & a_{2n} \\
\vdots & \vdots & \vdots & \ddots & \vdots\\
d_M &  a_{M1} & a_{M2} & \hdots & a_{Mn}\\
\end{block}
\end{blockarray}
\]

Now we can define the similarity between two documents $\{A,B\}$ as the  cosine 
of the angle $\theta$ between their vector representations:
\begin{equation}
\text{similarity} = 
\cos(\theta) = \frac{\mathbf{A} \cdot \mathbf{B}}{\|\mathbf{A}\| 
\|\mathbf{B}\|} = \frac{ \sum\limits_{i=1}^{n}{A_i  B_i} }{ 
\sqrt{\sum\limits_{i=1}^{n}{A_i^2}}  \sqrt{\sum\limits_{i=1}^{n}{B_i^2}} }
\label{eq:cos_sim}
\end{equation}

\subsection{Temporal association}
Once the similarity of two articles is calculated, their temporal association 
must 
be determined in order to consider the probability of the older article being 
the \emph{infector} of the other. In order to determine the most-likely 
\emph{infector} of 
an article, we ranked all articles by date of publications and looked within a 
fixed time window preceding the publication of each article, for the articles 
which are most semantically similar. The choice of the size of the time window 
was determined in order encompass the majority ($>95$\%) of previous similar 
articles (see figure \ref{fig:influence_range}).

\subsection{Reconstructing the Spread Network}
To reconstruct the spread network of the news, we defined the nodes of our 
network as the articles published on the subject chosen, and the edges as the 
infection events, i.e., for every article after the first one, it must have 
been influenced (infected)  by a previously published article. To qualify as an 
infector, an article must precede the infected article by less than 
$\gamma$ hours, and have a score of similarity (defined by 
(\ref{eq:cos_sim})) to 
the infected article of at least $\rho$.
The reconstruction 
procedure is summarized in the four steps below.

\begin{figure}[!ht]
 \centering
 \includegraphics[scale=0.7]{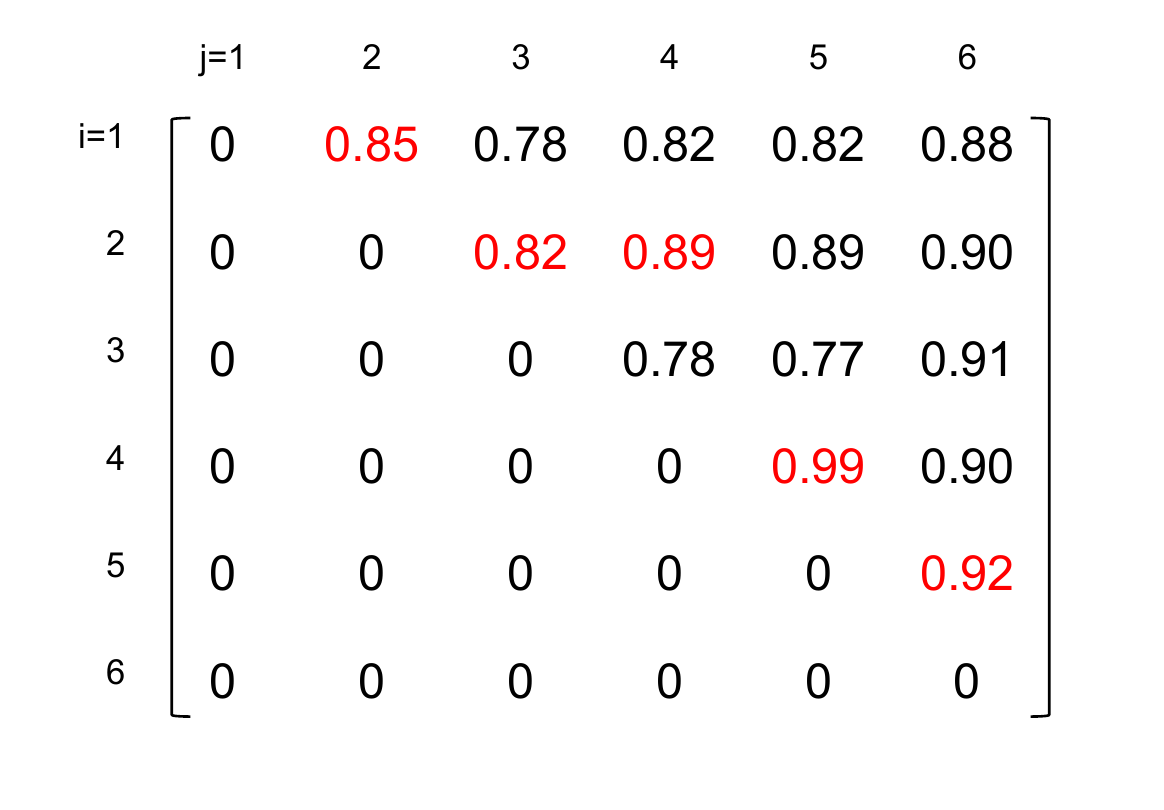}
 % sim_matrix.png: 445x311 pixel, 72dpi, 15.70x10.97 cm, bb=0 0 445 311
 \caption{This figure is a cut of our matrix $S$, it shows the first 6 articles.
 In red are the maximum similarity score for each column, which we use to define
 it's infector, per example, the article 4 has been infected by the
 article 2.}
 \label{fig:sim_matrix}
\end{figure}

\begin{enumerate}
\item Rank all articles in ascending publication time. Let $p_i$ denote the 
publication date of article $i$.
 \item Create upper triangular matrix $D$, where $d_{ij}={\cal 
H}(\gamma-\delta_{ij})*\delta_{ij}$ and $\delta_{ij}=p_j-p_i$. ${\cal H}$ is 
the 
Heaviside function.

 \item Create similarity matrix $S$. Where $s_{ij}$ is     
the similarity defined by equation (2) whenever $d_{ij}\neq0$ and $0$ 
otherwise.

 \item For each article $j$, we define its influencer $i$ as the article 
corresponding to $max(s_j)$ (see matrix \ref{fig:sim_matrix}).
\end{enumerate}

\subsection{Simulation model}
To test the hypothesis that news spread follows an epidemic process, we 
proposed an SIR model for the spread, following the formalism of 
\citet{pastor2015epidemic}. In this formalism, instead of modeling the status 
of a given individual  as Susceptible (S), Infectious (I) or Recovered (R), we 
model the probability of each article being in each of the states, in this 
case, an \textbf{S} article would be one which has yet to be published, an 
\textbf{I} one which is published and has been infected by the story and an 
\textbf{R} is one which is too old to influence new articles. This modeling 
leads us to equations \ref{eq:SIR}.

\begin{subequations}
\begin{empheq}[left=\empheqlbrace]{align}
\dfrac{d\rho^{I}_{i}}{dt} &= -\rho^{I}_{i}(t) + 
\lambda\rho_{i}^{S}(t)\sum_{j=1}^{N} a_{ij}\rho_{j}^{I}(t)\\
\dfrac{d\rho^{S}_{i}}{dt} &= - \lambda\rho_{i}^{S}(t) \sum_{j=1}^{N} 
a_{ij}\rho_{j}^{I}(t) 
\end{empheq}
\label{eq:SIR}
\end{subequations}

In equations \ref{eq:SIR}, $\rho^{I}_{i}(t)$ is the probability of 
article $i$ being in the infectious state at time $t$, similarly for 
$\rho^{S}_{i}(t)$; $a_{ij}$ is the probability of article $j$ being influenced 
by $i$ and comes from the adjacency matrix of the network. $\lambda$ is an 
adimensional transmission parameter given by $\lambda=\frac{\beta}{\mu}$. Time 
($t$) in these equations is also adimensional as it is scaled by $\mu$.

The network for the simulation is built from the same node set of the empirical 
data. The adjacency matrix $A$ is given by
\begin{equation}
      \begin{cases}
      i = j: a_{ij} = 0\\
      i \neq j: a_{ij} = \frac{N_{XY}}{N_Y}
      \end{cases}
 \label{eq:A}
    \end{equation}
    
where $N_{XY}$ is the number of times an article from publisher $X$ (the 
publisher of article $i$), has infected an article from publisher $Y$ (the 
publisher of article $j$) and $N_Y$ Is the total number of articles from 
publisher $Y$ that have been infected, regardless of publisher. These counts 
are derived from the empirical dataset.

The solution of this model generates the temporal dynamics of the probabilities 
described in (\ref{eq:SIR}). From the solutions, $\rho_{i}^{S}(t)$ and 
$\rho_{i}^{I}(t)$ we can derive realizations of states for each article, 
$S_i(t)$, $I_i(t)$, and $R_i(t)$.

To reconstruct the states, we must sample from the probability distribution 
the states at each time $t$, conditioning on the previous state. We follow the 
procedure:
\begin{enumerate}
 \item Let $S_t$, $I_t$ and $R_t$ be binary state vectors from article states 
at 
time $t$, where 1 means the article is in that state.
 \item Iterate from $t=0$ until the final time step available.
 \item For each time $t>0$ generate a newly infected $I_t^*$ vector, in which 
each element $i$ is a realization of a Bernoulli event with probability given 
by $\rho_{i}^{I}(t) \times S_{t-1}[i]$. 
\item Similarly to the previous step, sample a new $R_t$ vector, in which each 
element $i$ 
is a realization of a Bernoulli event with probability given by 
$\rho_{i}^{R}(t) \times I_{t-1}[i]$. 
\item Update $I_t =I_{t-1} - R_t + I_t^*$ 
\item Update $S_t=S_{t-1}- I_t^*$
\end{enumerate}

\subsection{Constructing the Simulated Spread Network}

From the state matrix $I$ we have which articles get infected at each time $t$. 
To create a spread network for the simulation we need to define the infectors 
for each
time. For that we used the probability matrix $A$ defined by equation 
(\ref{eq:A}). The following steps describe the entire procedure.

\begin{enumerate}
 \item Let $I_t$ be binary state vectors for articles at time t,
where 1 means the article is infected.
 \item Iterate from $t=1$ until the final time step available.
 \item For each article $i$ infected at $I_t$, obtain its probable infectors, 
$P_i$, by multiplying $I_{t-1}$ by the column $j$ of matrix $A$, where $j=i$.
 where the values are the probability of each article $j$ from $I_0$ has to 
infect $i$
 ($a_{ij}$ of the matrix A).
 \item Define the infector of $i$ by sampling from a multinomial distribution 
with $p=P_i$.
\end{enumerate}

The figure \ref{fig:sim_net} shows the procedure:

\begin{figure}[h]
 \centering
 \includegraphics[scale=0.4]{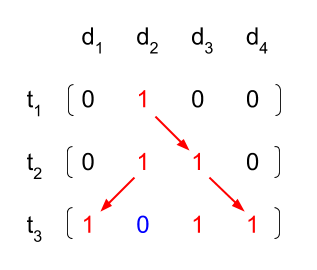}
 % sim_network.png: 317x270 pixel, 72dpi, 11.18x9.52 cm, bb=0 0 317 270
 \caption{The arrows indicates the infector for each article. The red articles
 are the ones infected, that is, the ones that can spread the infection, and the 
blues 
 are the ones that had recovered.}
 \label{fig:sim_net}
\end{figure}

\section{Results}

\begin{figure}[!ht]
 \centering
 \includegraphics[width=\textwidth]{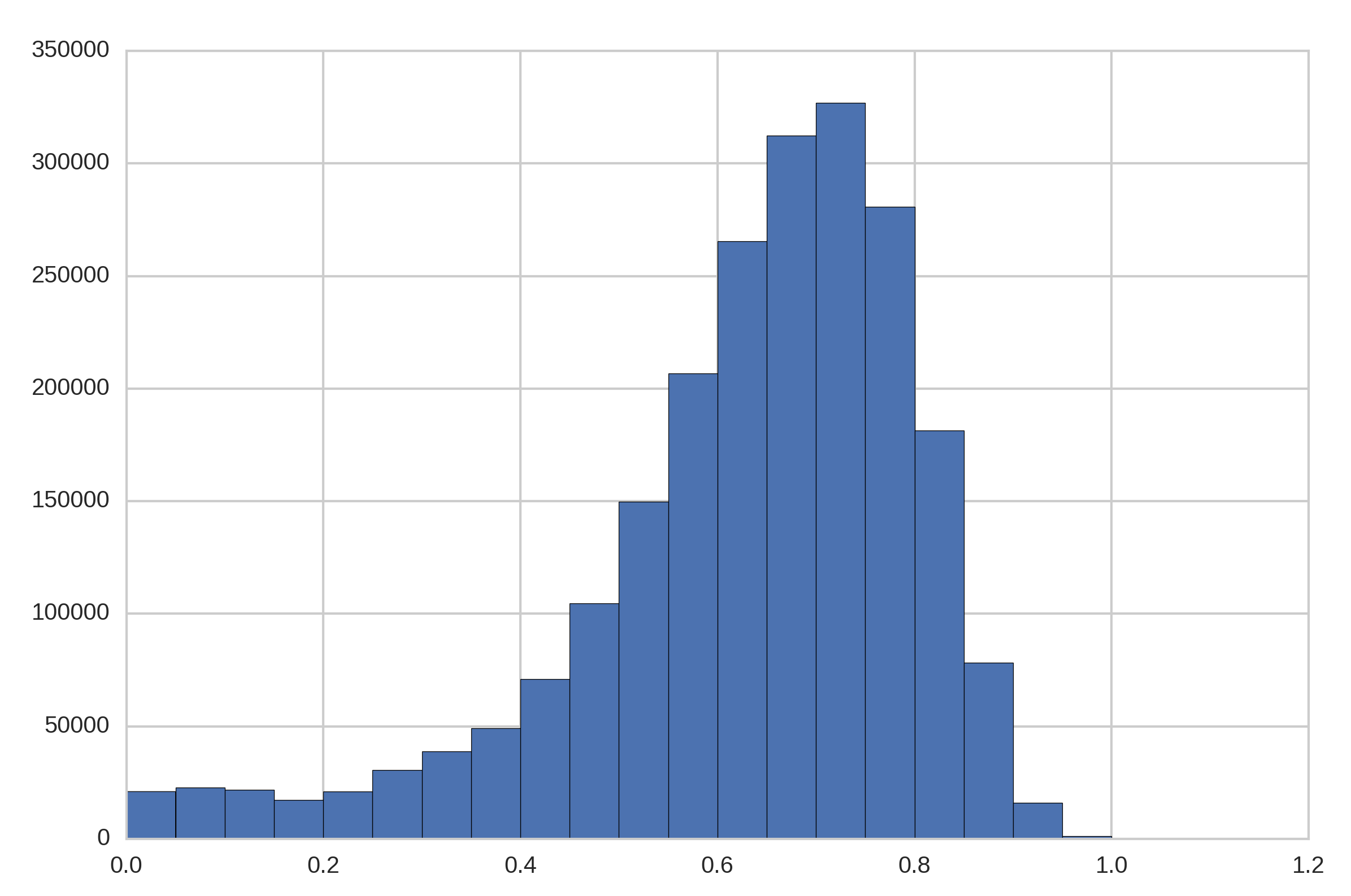}
 % similarity_dist.png: 0x0 pixel, 300dpi, 0.00x0.00 cm, bb=
 \caption{Distribution of the pairwise similarities among all articles.}
 \label{fig:sim_pair}
\end{figure}

The dataset used is the result of a very specific search on a news 
articles database, therefore we can expect to the articles to display a great 
similarity among themselves. Figure \ref{fig:sim_pair}, shows the distribution 
of pairwise similarities that were used to construct the empirical influence 
network. 

\begin{figure}[!ht]
 \centering
 \includegraphics[width=\textwidth]{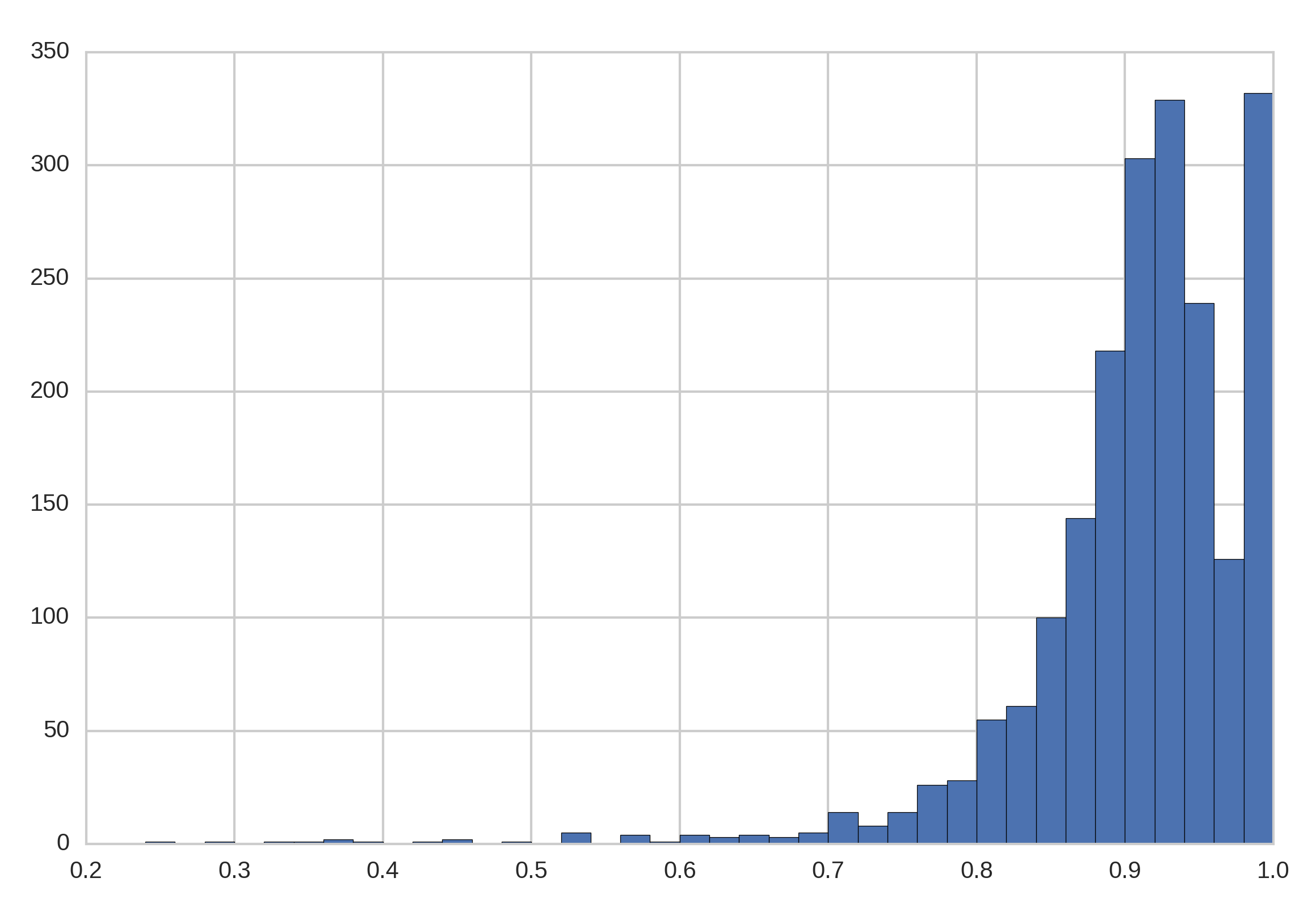}
 % similarity_dist_most_similar.png: 432x288 pixel, 72dpi, 15.24x10.16 cm, bb=0 
0 432 288
 \caption{Distribution of similarities of the most similar article to each 
article in the collection.}
 \label{fig:sim_most}
\end{figure}

When we look to the most similar pair for each article we can notice that for 
almost 
every article there is at least one other with similarity equal or greater than 
0.8, 
as we can see in figure \ref{fig:sim_most}.
Identical articles (similarity equals to 1) were not considered for edge 
formation.

\begin{figure}[!ht]
 \centering
 \includegraphics[width=\textwidth]{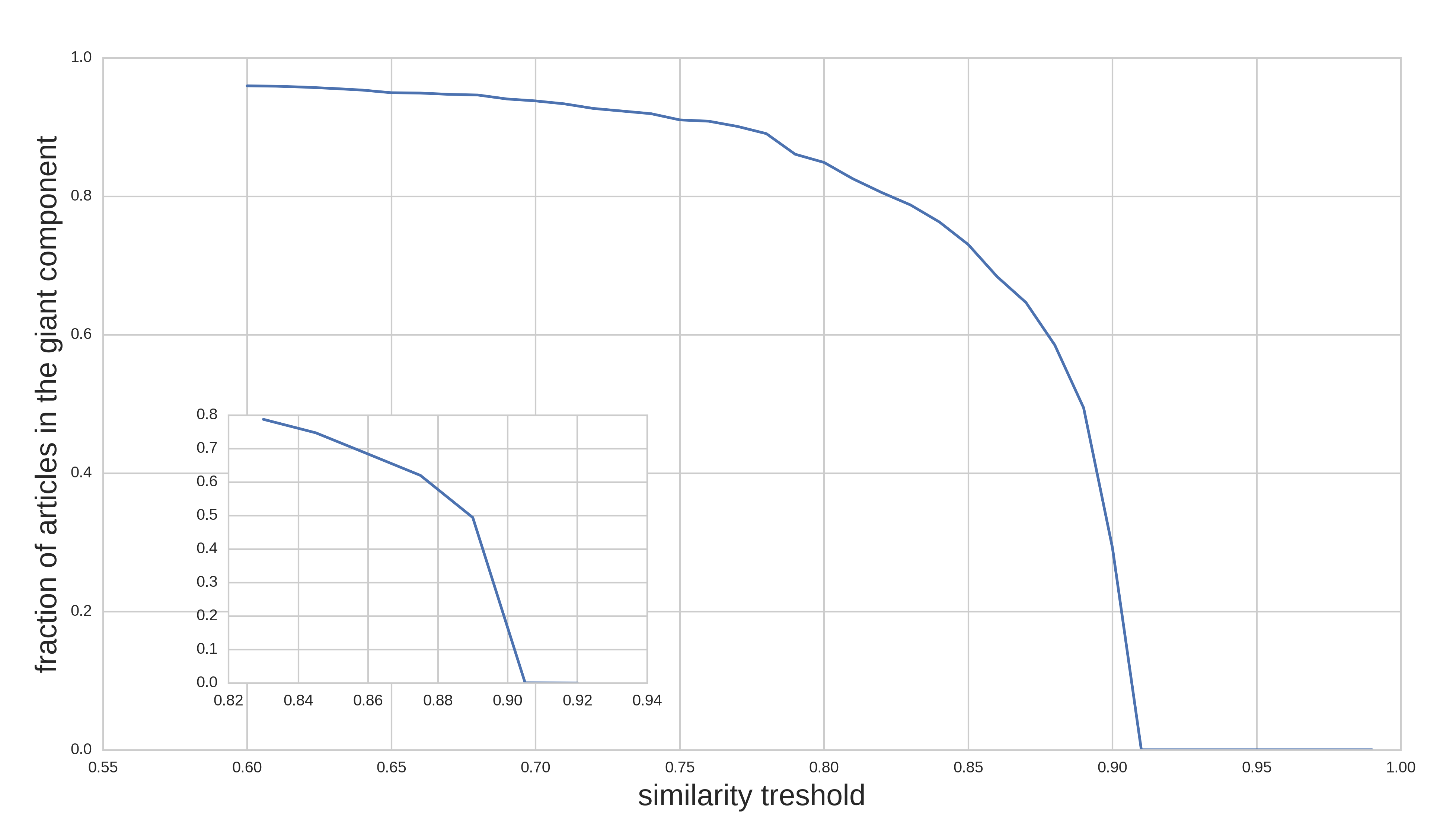}
 % similarity.png: 0x0 pixel, 300dpi, 0.00x0.00 cm, bb=
 \caption{Similarity threshold for the reconstruction of 
the influence network.}
 \label{fig:sim_thresh}
\end{figure}

Figure \ref{fig:sim_thresh} shows the similarity threshold of the influence 
network.
In order to have a giant component in the network that contains at least $80\%$ 
of our
articles, we need to consider a minimum of $0.8$ score similarity. Therefore, we
defined $\rho=0.8$.

\begin{figure}[!ht]
 \centering
 \includegraphics[width=\textwidth]{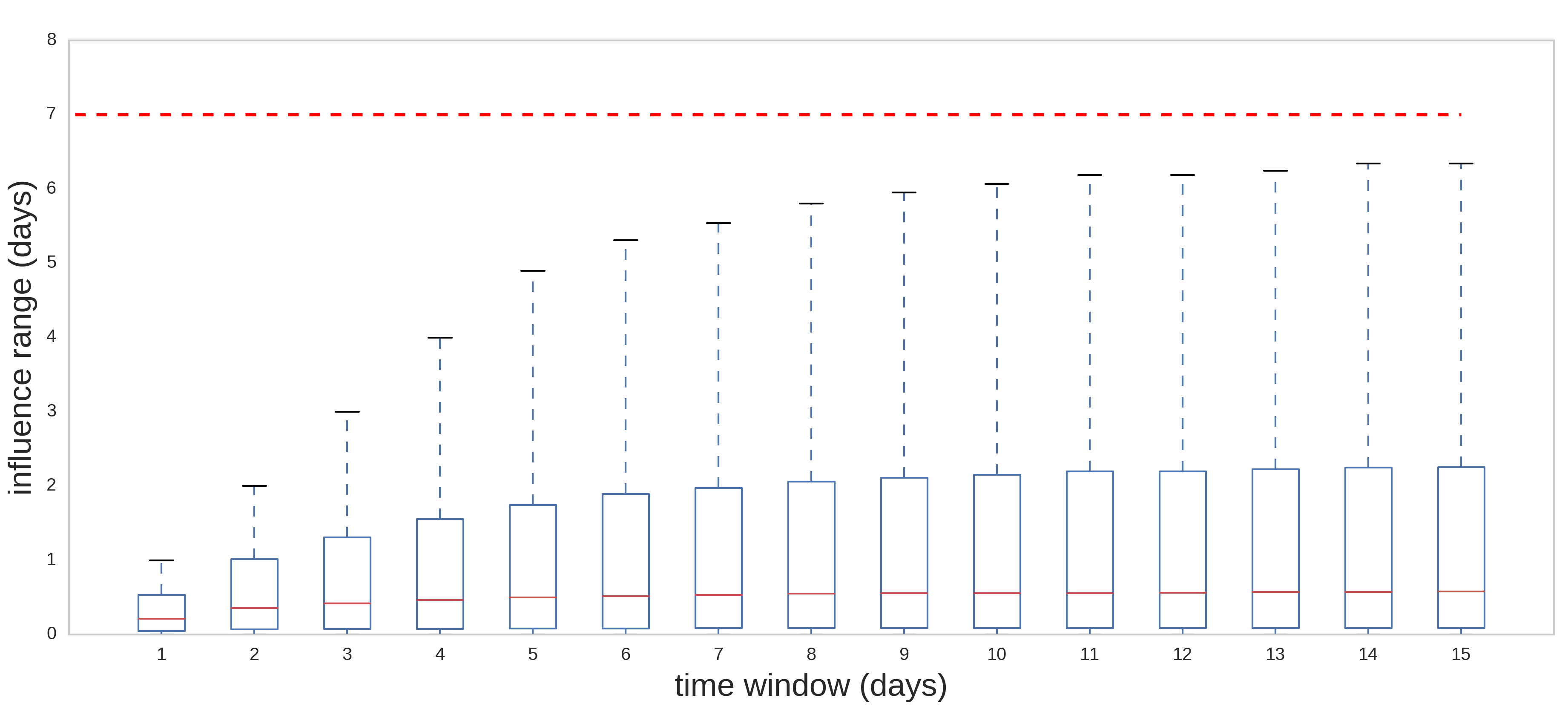}
 % influence_range.png: 0x0 pixel, 300dpi, 0.00x0.00 cm, bb=
 \caption{Distribution of time lag from influencer for multiple time window 
lengths. Notice that no article lags more than seven days from its influencer.}
 \label{fig:influence_range}
\end{figure}

To determine the optimal time window $\gamma$ in which to search for 
influencers, we 
looked at the distribution of time lags from the most similar article 
(most likely influencer) at various window lengths (figure 
\ref{fig:influence_range}). 
Even for time windows as long as 15 days, 95\% of the influencers where within 
7 
days of the articles they influenced.

\begin{figure}[!ht]
 \centering
 \includegraphics[width=\textwidth]{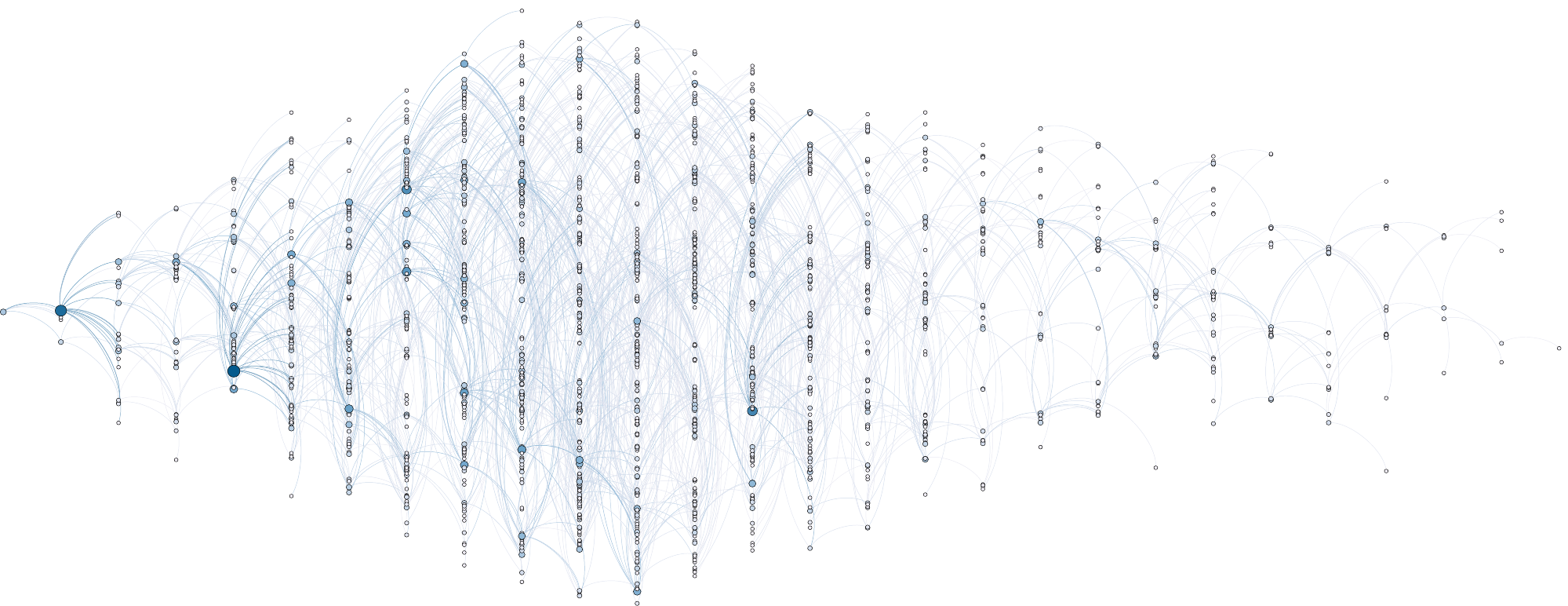}
 % empirical_graph.pdf: 595x842 pixel, 72dpi, 20.99x29.70 cm, bb=0 0 595 842
 \caption{The nodes are scaled by the out-degree and are arranged by generation. 
i. e., in the first column is
 the node that was first published, the second one are those influenced by him 
and so on.}
 \label{fig:net}
\end{figure}

To create the spread network, we defined influence based on the time lag from 
each pair of articles and from their similarity. Following the previous 
analysis,
we defined $\rho=0.8$ and $\gamma=168$, that means, the infector must preceded 
the 
infected article by less than 168 hours (7 days) and have at least a 0.8 score 
of 
similarity. We can see in figure
\ref{fig:net} how our network looks like. 

\subsection{Simulation}

\begin{figure}[!ht]
 \centering
 \includegraphics[width=\textwidth]{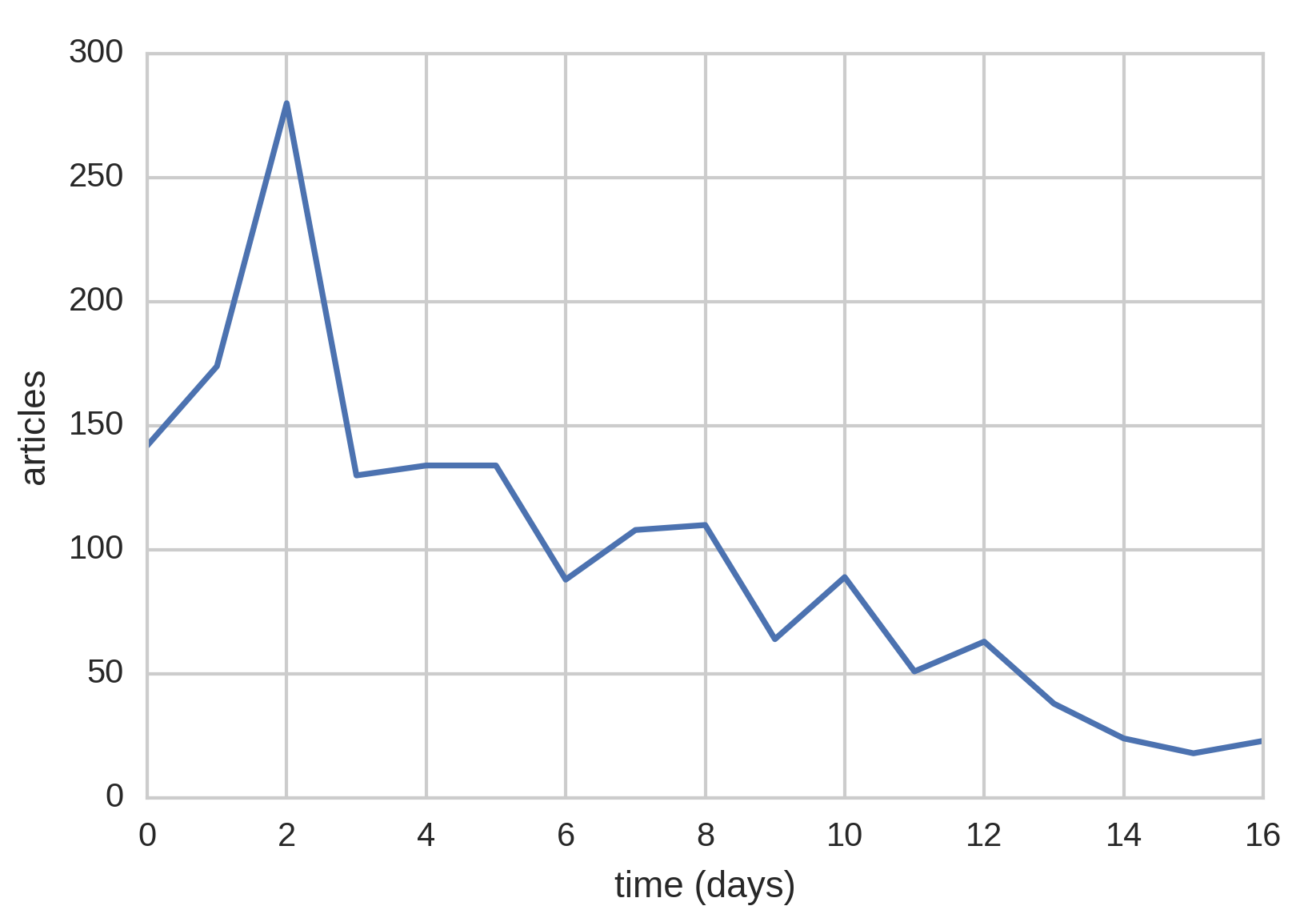}
 % data_dist.png: 1619x1125 pixel, 72dpi, 57.11x39.69 cm, bb=0 0 1619 1125
 \caption{Number of articles published per day}
 \label{fig:data_dist}
\end{figure}

\begin{figure}[!ht]
 \centering
 \includegraphics[width=\textwidth]{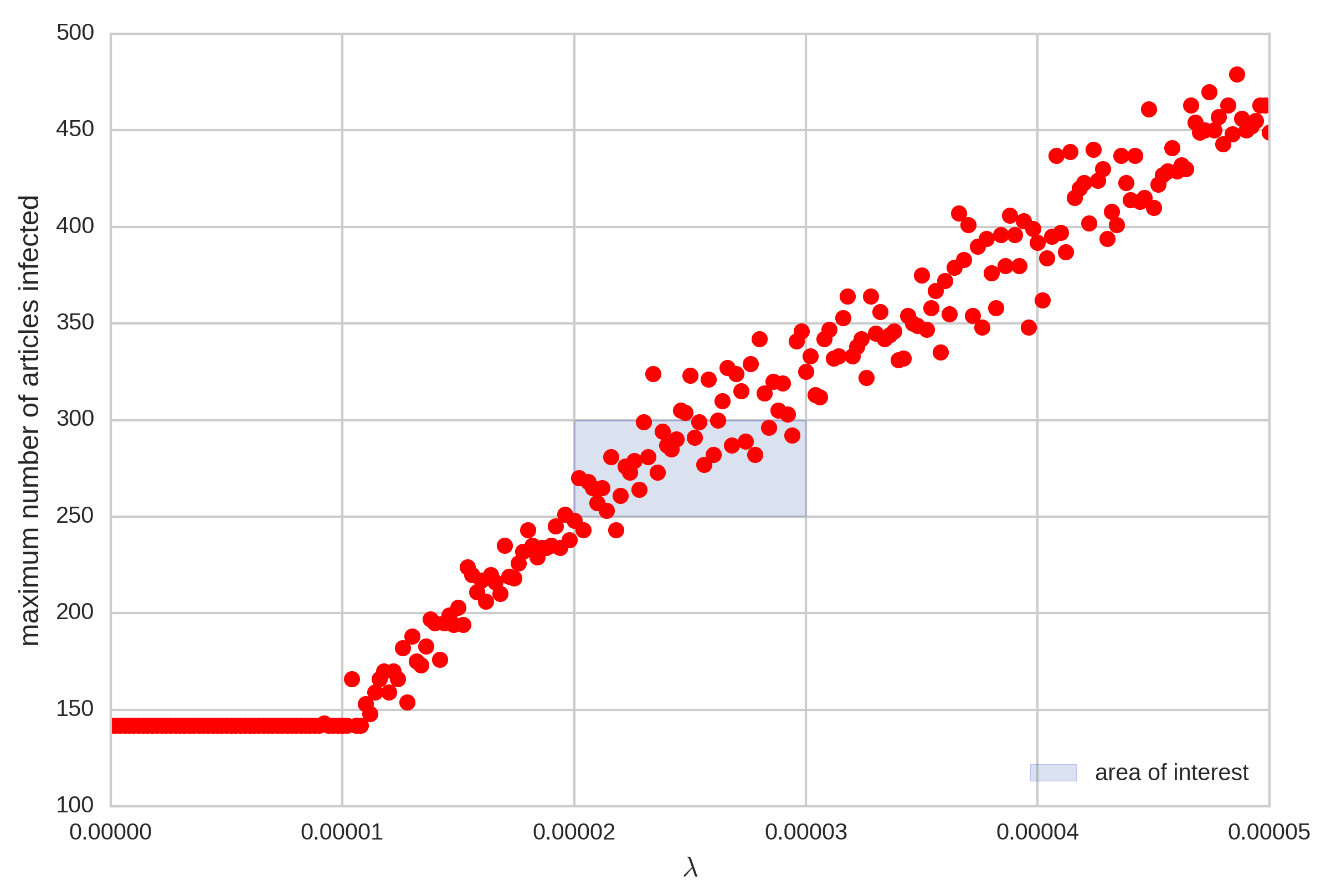}
 % lambda_validation.png: 2700x1800 pixel, 300dpi, 22.86x15.24 cm, bb=0 0 648 
432
 \label{fig:lambda}
 \caption{Total number of articles infected between $0<\lambda<0.00005$. The 
blue area
 is the area where the peak of the simulation is the same as the peak of the 
dataset
 distribution, threfore is the area where the $\lambda$ values were tested for 
our
 simulation.}
\end{figure}

Looking at the publication date distribution (figure \ref{fig:data_dist}) we 
notice that the maximum number 
of articles published in a day was between 250 and 300. We derive the simulation
parameters from this distribution. For example, on figure \ref{fig:lambda} we 
plot
the peak of the infection for a range($[0,0.00005]$) of $\lambda$ values. From 
that distribution 
of peak magnitudes we selected a lambda to match the empirical peak: 
$0.00002<\lambda<0.00003$. 

\begin{figure}[!ht]
 \centering
 \includegraphics[width=\textwidth]{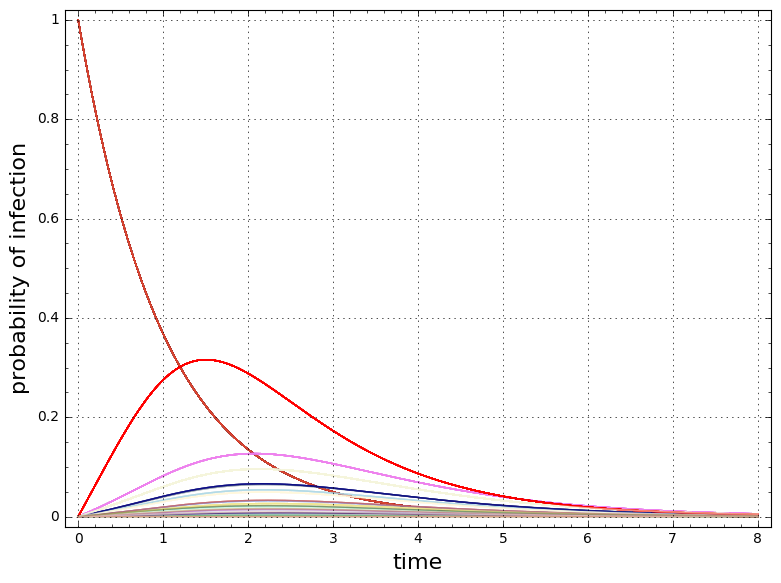}
 % simulation.png: 781x584 pixel, 100dpi, 19.84x14.83 cm, bb=0 0 562 420
 \caption{Simulation for $\lambda = 0.0000215$. Each curve represents the 
n the infectious state as a function of time, for every article. The time units 
are $1/\mu$.}
 \label{fig:simu}
\end{figure}

\begin{figure}[!ht]
 \centering
 \includegraphics[width=\textwidth]{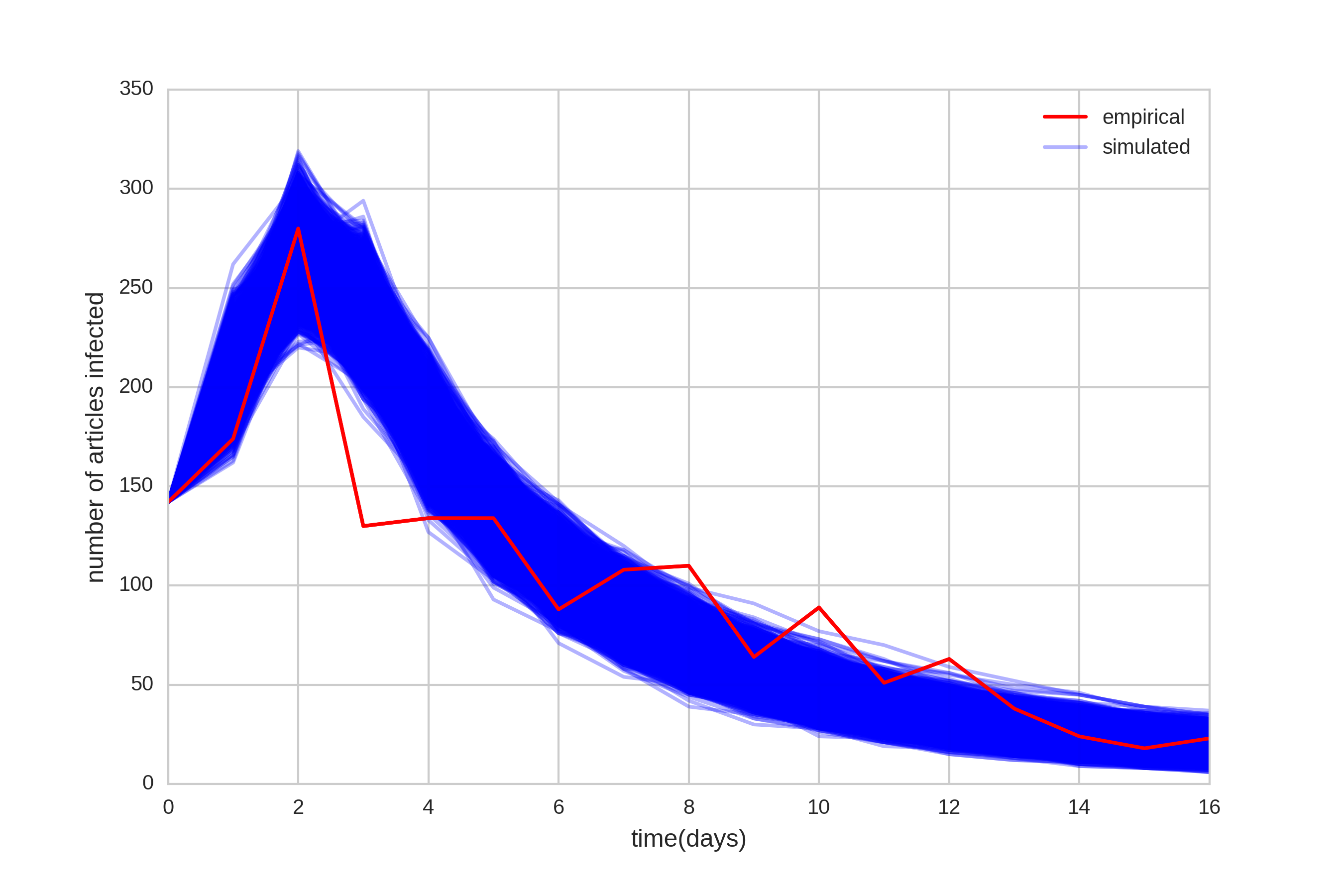}
 % sim_compar.png: 432x288 pixel, 72dpi, 15.24x10.16 cm, bb=0 0 432 288
 \caption{The blue curves are the 10.000 realizations of the state matrix. Notice that the simulated dynamics match the empirical curve.}
 \label{fig:sim_emp}
\end{figure}

From the simulation (figure \ref{fig:simu}) we obtain the state matrix, which we 
use to
compare the simulated infection distribution with the original data. Then we ran 
10 thousand simulations 
to show that the model proposed matches the real world dynamics of news articles 
influences
(figure \ref{fig:sim_emp}).

\section{Discussion}

In this paper, we presented a methodology for reconstructing the network representing the spread of news in digital media. The results proposed started  from a well defined subset of articles with high semantic similarities. However we believe the criteria of similarity used to reconstruct the network would work even on a random sample or articles, provided that it was large enough to contain a good portion of the putative spread network one is trying to characterize. In other words, the reconstruction algorithm can be used to detect contagious structures within any large enough collection of news articles. 

We also demonstrated that a classical SIR process over the network is driving the spread dynamics. This means that the if one is able to observe the start of the spread, the overall reach and time of persistence in the media can be predicted from analytical results available for the SIR model.

The news subject selected, "Charlie Hebdo attack", represents a very spontaneous media coverage given the great surprise with which it happened, but also due the homogeneous response of the global media condemning the cowardly attack. We believe that deviations from the classical SIR dynamics shown here can hint at some form of media manipulation, but that hypothesis remains to be tested based on well defined cases, such as purchased media coverage during political campaigns, etc. With the current concerns about ``fake" media pieces~\citep{Berkowitz_Schwartz_2016}, Perhaps the methods presented here can help to discriminate authentic media articles from fake ones based on their spread dynamics or influence patterns. We already began to see some attempts to automatically detect fake news~\citep{Jin_Cao_Zhang_Zhou_Tian_2016,Rubin_Conroy_Chen_Cornwell}, but they mostly rely on linguistic cues. We believe that qualitative and quantitative aspects of the spread networks can also be of use.



%% References

\newpage

\bibliographystyle{model1-num-names}

\begin{thebibliography}{10}
\expandafter\ifx\csname natexlab\endcsname\relax\def\natexlab#1{#1}\fi
\providecommand{\bibinfo}[2]{#2}
\ifx\xfnm\relax \def\xfnm[#1]{\unskip,\space#1}\fi
%Type = Article
\bibitem[{Hermida et~al.(2012)Hermida, Fletcher, Korell, and
  Logan}]{hermida2012share}
\bibinfo{author}{A.~Hermida}, \bibinfo{author}{F.~Fletcher},
  \bibinfo{author}{D.~Korell}, \bibinfo{author}{D.~Logan},
\newblock \bibinfo{title}{Share, like, recommend: Decoding the social media
  news consumer},
\newblock \bibinfo{journal}{Journalism Studies} \bibinfo{volume}{13}
  (\bibinfo{year}{2012}) \bibinfo{pages}{815--824}.
%Type = Incollection
\bibitem[{Romero et~al.(2011)Romero, Galuba, Asur, and
  Huberman}]{romero2011influence}
\bibinfo{author}{D.~M. Romero}, \bibinfo{author}{W.~Galuba},
  \bibinfo{author}{S.~Asur}, \bibinfo{author}{B.~A. Huberman},
\newblock \bibinfo{title}{Influence and passivity in social media},
\newblock in: \bibinfo{booktitle}{Machine learning and knowledge discovery in
  databases}, \bibinfo{publisher}{Springer}, \bibinfo{year}{2011}, pp.
  \bibinfo{pages}{18--33}.
%Type = Article
\bibitem[{Pastor-Satorras et~al.(2015)Pastor-Satorras, Castellano, Van~Mieghem,
  and Vespignani}]{pastor2015epidemic}
\bibinfo{author}{R.~Pastor-Satorras}, \bibinfo{author}{C.~Castellano},
  \bibinfo{author}{P.~Van~Mieghem}, \bibinfo{author}{A.~Vespignani},
\newblock \bibinfo{title}{Epidemic processes in complex networks},
\newblock \bibinfo{journal}{Reviews of modern physics} \bibinfo{volume}{87}
  (\bibinfo{year}{2015}) \bibinfo{pages}{925}.
%Type = Article
\bibitem[{Bettencourt et~al.(2006)Bettencourt, Cintr{\'o}n-Arias, Kaiser, and
  Castillo-Ch{\'a}vez}]{bettencourt2006power}
\bibinfo{author}{L.~M. Bettencourt}, \bibinfo{author}{A.~Cintr{\'o}n-Arias},
  \bibinfo{author}{D.~I. Kaiser}, \bibinfo{author}{C.~Castillo-Ch{\'a}vez},
\newblock \bibinfo{title}{The power of a good idea: Quantitative modeling of
  the spread of ideas from epidemiological models},
\newblock \bibinfo{journal}{Physica A: Statistical Mechanics and its
  Applications} \bibinfo{volume}{364} (\bibinfo{year}{2006})
  \bibinfo{pages}{513--536}.
%Type = Inproceedings
\bibitem[{Mihalcea et~al.(2006)Mihalcea, Corley, and
  Strapparava}]{mihalcea2006corpus}
\bibinfo{author}{R.~Mihalcea}, \bibinfo{author}{C.~Corley},
  \bibinfo{author}{C.~Strapparava},
\newblock \bibinfo{title}{Corpus-based and knowledge-based measures of text
  semantic similarity},
\newblock in: \bibinfo{booktitle}{AAAI}, volume~\bibinfo{volume}{6}, pp.
  \bibinfo{pages}{775--780}.
%Type = Article
\bibitem[{Mikolov et~al.(2013)Mikolov, Chen, Corrado, and
  Dean}]{mikolov2013efficient}
\bibinfo{author}{T.~Mikolov}, \bibinfo{author}{K.~Chen},
  \bibinfo{author}{G.~Corrado}, \bibinfo{author}{J.~Dean},
\newblock \bibinfo{title}{Efficient estimation of word representations in
  vector space},
\newblock \bibinfo{journal}{arXiv preprint arXiv:1301.3781}
  (\bibinfo{year}{2013}).
%Type = Article
\bibitem[{Hiemstra(2000)}]{Hiemstra_2000}
\bibinfo{author}{D.~Hiemstra},
\newblock \bibinfo{title}{A probabilistic justification for using tf$\times$
  idf term weighting in information retrieval},
\newblock \bibinfo{journal}{International Journal on Digital Libraries}
  \bibinfo{volume}{3} (\bibinfo{year}{2000}) \bibinfo{pages}{131–139}.
%Type = Article
\bibitem[{Berkowitz and Schwartz(2016)}]{Berkowitz_Schwartz_2016}
\bibinfo{author}{D.~Berkowitz}, \bibinfo{author}{D.~A. Schwartz},
\newblock \bibinfo{title}{Miley, cnn and the onion: When fake news becomes
  realer than real},
\newblock \bibinfo{journal}{Journalism Practice} \bibinfo{volume}{10}
  (\bibinfo{year}{2016}) \bibinfo{pages}{1–17}.
%Type = Article
\bibitem[{Jin et~al.(2016)Jin, Cao, Zhang, Zhou, and
  Tian}]{Jin_Cao_Zhang_Zhou_Tian_2016}
\bibinfo{author}{Z.~Jin}, \bibinfo{author}{J.~Cao}, \bibinfo{author}{Y.~Zhang},
  \bibinfo{author}{J.~Zhou}, \bibinfo{author}{Q.~Tian},
\newblock \bibinfo{title}{Novel visual and statistical image features for
  microblogs news verification},
\newblock \bibinfo{journal}{IEEE Transactions on Multimedia}
  (\bibinfo{year}{2016}).
%Type = Article
\bibitem[{Rubin et~al.(????)Rubin, Conroy, Chen, and
  Cornwell}]{Rubin_Conroy_Chen_Cornwell}
\bibinfo{author}{V.~L. Rubin}, \bibinfo{author}{N.~J. Conroy},
  \bibinfo{author}{Y.~Chen}, \bibinfo{author}{S.~Cornwell},
\newblock \bibinfo{title}{Fake news or truth? using satirical cues to detect
  potentially misleading news.}  (????).

\end{thebibliography}






% \end{thebibliography}

\end{document}